\documentclass[twocolumn,preprintnumbers,superscriptaddress,nofootinbib,aps,prd,floatfix]{revtex4}
\usepackage{footmisc,enumerate}
\usepackage{subfigure}
\usepackage{amsmath,amssymb}
\usepackage{graphicx}
\usepackage{bbm,slashed}
\usepackage{xspace,slashed}
\usepackage{hyperref}
\hypersetup{colorlinks=true, citecolor=blue, urlcolor=blue, linkcolor=blue}
\usepackage[normalem]{ulem}

\newcommand{\lam}{\lambda}

\begin{document}

\providecommand{\abs}[1]{\lvert#1\rvert}

\newcommand{\Znunujets}{(Z\to{\nu\bar{\nu}})+\text{jets}}
\newcommand{\Welnujets}{(W\to{\ell\nu})+\text{jets}}
\newcommand{\Znunujet}{(Z\to{\nu\bar{\nu}})+\text{jet}}
\newcommand{\Welnujet}{(W\to{\ell\nu})+\text{jet}}

\title{On interference effects in top-philic decay chains}
\begin{abstract}
Sizeable exotic Higgs boson production through gluon fusion via top quark loops is correlated with large Higgs decay probabilities into top quark final states, if these are kinematically accessible. It is known that $gg\to S \to t\bar t$ is particularly susceptible to signal-background interference effects that can significantly impact discovery sensitivities. In such an instance, identifying more robust signatures to enhance the discovery sensitivity becomes necessary, shifting phenomenological focus to final states that show a reduced destructive signal-background interference. In this work, we discuss the phenomenological relevance of signal-signal and signal-background interference effects for decay chains. In particular, we identify asymmetric cascade decays in models of scalar extensions of the SM Higgs sector as new promising candidates. In parallel, characteristic interference patterns can provide a formidable tool for indirect CP spectroscopy of BSM sectors if a discovery is made in the future. This motivates associated searches at present and future colliders as robust discovery modes of new physics, in addition to serving as a sensitive tool for the reconstruction of the underlying UV electroweak potential.
\end{abstract}

\author{Oliver Atkinson}\email{o.atkinson.1@research.gla.ac.uk} 
\affiliation{SUPA, School of Physics \& Astronomy, University of Glasgow, Glasgow G12 8QQ, UK\\[0.1cm]}
\author{Christoph Englert} \email{christoph.englert@glasgow.ac.uk}
\affiliation{SUPA, School of Physics \& Astronomy, University of Glasgow, Glasgow G12 8QQ, UK\\[0.1cm]}
\author{Panagiotis Stylianou}\email{p.stylianou.1@research.gla.ac.uk} 
\affiliation{SUPA, School of Physics \& Astronomy, University of Glasgow, Glasgow G12 8QQ, UK\\[0.1cm]}


\maketitle

\section{Introduction}
\label{sec:intro}
The search for new physics beyond the Standard Model (BSM) is one of the highest
priorities of the Large Hadron Collider (LHC) experiments. While astrophysical observations
highlight the need to extend the SM to incorporate additional sources of CP violation or dark matter candidates,
current searches for new physics have not revealed the presence of BSM phenomena in a
statistically significant way. 

On the one hand, one can interpret the consistency of
experimental findings with the SM expectations as an indication of a large gap between the LHC-relevant
TeV scale and the new scale of BSM interactions, which motivates the application
of effective field theory techniques to classify deviations from the SM in a theoretically
consistent way~\cite{Weinberg:1978kz,Buchmuller:1985jz,Burges:1983zg,Leung:1984ni,Hagiwara:1986vm,Grzadkowski:2010es,Dedes:2017zog}
(for a review see~\cite{Brivio:2017vri}). 

On the other hand, if the scale of new physics is too low to meet the validity 
assumptions of EFT modifications (alongside their perturbative matching to high-scale UV completions),
SM consistency of LHC data should be interpreted in terms of constraints for concrete new physics models. Particularly
motivated in this context are Higgs sector extensions as such models can address shortcomings
of the SM, ranging from additional CP violation, over strong first order phase transitions, to the relation of the TeV scale with dark matter. Immediate phenomenological consequences of such scenarios are new resonant structures in top final states that straightforwardly arise as a consequence of custodial isospin singlet mixing present in an extended Higgs sector. If production of new, heavy scalar states is significant via gluon fusion $gg\to S$ (as found in multi-singlet, doublet and triplet Higgs extensions), the decay $S\to t\bar t$ is motivated as a search strategy, especially when new states $S$ are gauge-phobic (e.g. the CP-odd Higgs boson in two-Higgs-doublet \hbox{models}).

It is known~\cite{Gaemers:1984sj,Dicus:1994bm,Bernreuther:1998qv,Jung:2015gta,Frederix:2007gi,Barger:2006hm,Craig:2015jba,Bernreuther:2015fts,Carena:2016npr,Hespel:2016qaf,BuarqueFranzosi:2017qlm,BuarqueFranzosi:2017jrj,Djouadi:2019cbm} that this avenue can be significantly impacted by interference of new physics with 
the competing irreducible SM background contributions, significantly impacting the exclusion and discovery potential of such analyses. In particular, as shown in Ref.~\cite{Basler:2019nas}, such interference effects
become relevant for Higgs sector extensions with parameter choices that are consistent with current experimental 
constraints. 

The net effect of such interference is twofold. Firstly, rate expectations on the basis of a combination of production
cross section and decay probability (i.e. the application of the narrow width approximation) become invalid. From a theoretical point of view, this is troubling as high-precision results for Higgs production are not directly applicable, and the ad-hoc use of Breit-Wigner propagators as an inappropriate approximation to the consistent inclusion of unstable particles can a have additional sizeable effects~\cite{Seymour:1995np,Papavassiliou:1997fn,Papavassiliou:1996zn,Gambino:1999ai,Grassi:2001bz,Goria:2011wa}. Secondly, and perhaps experimentally more importantly, techniques that target a resonance structure overestimate the sensitivity to new physics
signals unless they include model-dependent corrections to the limit setting~\cite{Aaboud:2017hnm,Sirunyan:2019wph}.

The possibility of using the information that is provided through the characteristic distortion of the BSM resonances' Breit-Wigner distribution has been analysed before~\cite{Carena:2016npr,Djouadi:2019cbm} (see also \cite{Dixon:2003yb,Martin:2012xc,Dixon:2013haa,Martin:2013ula,Englert:2015zra}). In this work we pursue a different avenue, with the aim of regaining sensitivity to new states when interference is a particularly limiting factor effect for a physics discovery. We will demonstrate that in such an instance, asymmetric cascade decay chains with different intrinsic mass scales, as investigated recently in Ref.~\cite{Robens:2019kga,Englert:2020ntw}, are less impacted by destructive interference, thus providing a robust tool for BSM investigations, complementary to the standard searches that are currently pursued at the LHC (e.g.~\cite{Aaboud:2017hnm,Sirunyan:2019wph}).

We organise this paper as follows: we first analyse signal-background interference using a simplified model approach in Sec.~\ref{sec:simp} to highlight the importance of CP phases for expected interference effects in asymmetric cascade decays and decay chains. In Sec.~\ref{sec:tsing} we contextualise these results with a detailed analysis of signal-signal and signal-background interference in the two-singlet extended SM; see the recent~\cite{Robens:2019kga} for a detailed discussion. In this scenario, the decay $H_3\to H_2 h$ with $m_{H_3}> m_{H_2} > m_h \simeq 125~\text{GeV}$ can be sizeable, while $H_{3,2}\to t\bar t$ can be severely impacted by interference effects, so that $pp\to H_3 \to (H_2 \to t\bar t) (h\to b\bar b)$  offers a promising (yet challenging) phenomenological avenue, see~\cite{Englert:2020ntw}. We also comment on the impact of these effects for a 3/ab extrapolation of the LHC in Sec.~\ref{sec:tsing} for parameter choices where the cascade decay can be observed with large statistical significance. We conclude in Sec.~\ref{sec:conc}.

\section{Cascade interference: A simplified model analysis}
\label{sec:simp}
To enhance the SM with richer scalar phenomenology, we introduce two additional physical scalar degrees of freedom $H_2$, $H_3$ via the simplified Lagrangian 
\begin{multline}
\label{eq:frlagrangian}
	{\cal{L}} = {\cal{L}}_{\text{SM}} + \\\sum_{i=2,3} \frac{H_i}{v} \left[ C_{H_i}^{\text{even}} \frac{g_s^2}{16 \pi^2} G_{\mu\nu}^{a} G^{a\mu\nu} + C_{H_i}^{\text{odd}} \frac{g_s^2}{16 \pi^2} G_{\mu\nu}^{a} \widetilde{G}^{a\mu\nu}\right] \\ -  \frac{\lam_{H_2 h h}}{2} H_2 h^2 - \frac{\lam_{H_3 h h}}{2}H_3 h^2 - \lam_{H_3 H_2 h} H_3 H_2 h \\ - y^E_{H_3} \frac{m_t}{v} H_3 \bar{t}_L t_R - y^E_{H_2} \frac{m_t}{v} H_2 \bar{t}_L t_R + \text{h.c.}\ \\- i y^O_{H_3} {m_t \over v}H_3 \bar{t}_L t_R - i y^O_{H_2}  {m_t \over v}  H_2 \bar{t}_L t_R   + \text{h.c.}\;.
\end{multline}
We consider trilinear couplings $\lam_{ijk}$ ($i,j,k \in \{ h, H_2, H_3\}$), introducing interactions between the BSM scalars and the SM Higgs $h$ and allowing the possibility of detecting a cascade decay signal in colliders. The $ggH_i$ contact interactions capture the full top mass dependence by performing a one-loop matching calculation~\cite{Plehn:2009nd,Dawson:2018dcd}. We will limit ourselves to the region $m_{H_3}>m_{H_2}>2m_t$; choices $y^{E/O}_{H_i}=1$ refer to a CP-even/odd SM-like Higgs boson.

\begin{figure}[!t]
\subfigure[\label{fig:feyndiaga}]{\includegraphics[width=0.28\textwidth]{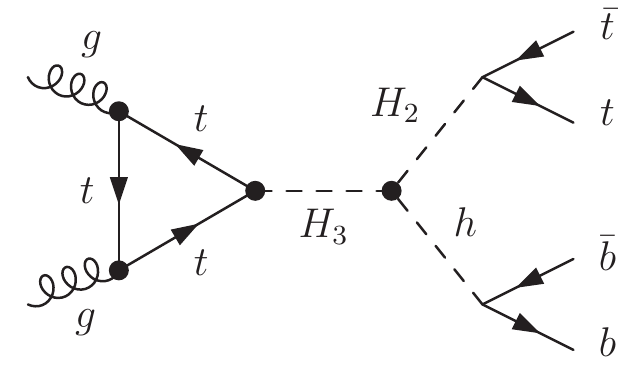}}\\
\subfigure[\label{fig:feyndiagb}]{\includegraphics[width=0.28\textwidth]{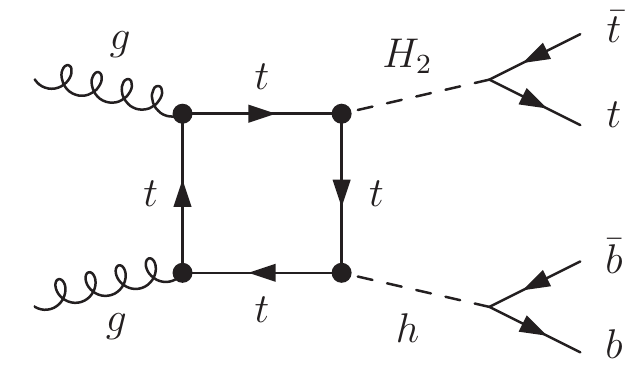}}
\caption{Representative Feynman diagrams contributing to the resonant (a) and non-resonant BSM signal (b). $H_3$ on-shell signal-signal interference arises from interference between diagrams (a) and (b). Not depicted are background topologies that contribute to $t\bar t b\bar b$ production in the SM. For details on implementation and evaluation see the text.\label{fig:feyndiag}}
\end{figure}

\subsection{Event generation for signal-background interference}
We model the signal of the scalar extended SM Lagrangian in Eq.~\ref{eq:frlagrangian} using {\sc{FeynRules}}~\cite{Christensen:2008py,Alloul:2013bka} and saving the interaction rules in the {\sc{UFO}}~\cite{Degrande:2011ua} format. The events are then generated with {\sc{MadEvent}}~\cite{Alwall:2011uj,deAquino:2011ub,Alwall:2014hca}, modified to include the full top-loop matching via modified {\sc{Helas}} routines.  We have cross-checked this implementation analytically, and numerically against the results of Refs.~\cite{Spira:1995mt,Anastasiou:2009kn}. 

In this discussion, we focus on the interference of all contributing SM backgrounds with the asymmetric cascade decay signal to highlight the different phenomenological outcomes that we can expect due to the presence of potentially sizeable interference contributions. In general these will be model-dependent, in particular because signal-signal interference can be a sizeable effect~\cite{Basler:2019nas}. We will return to this question, when we consider the concrete scenario of Sec.~\ref{sec:tsing}.

\begin{figure*}[!p]
\begin{center}
	\subfigure[\label{fig_2a}]{\includegraphics[width=0.48\textwidth]{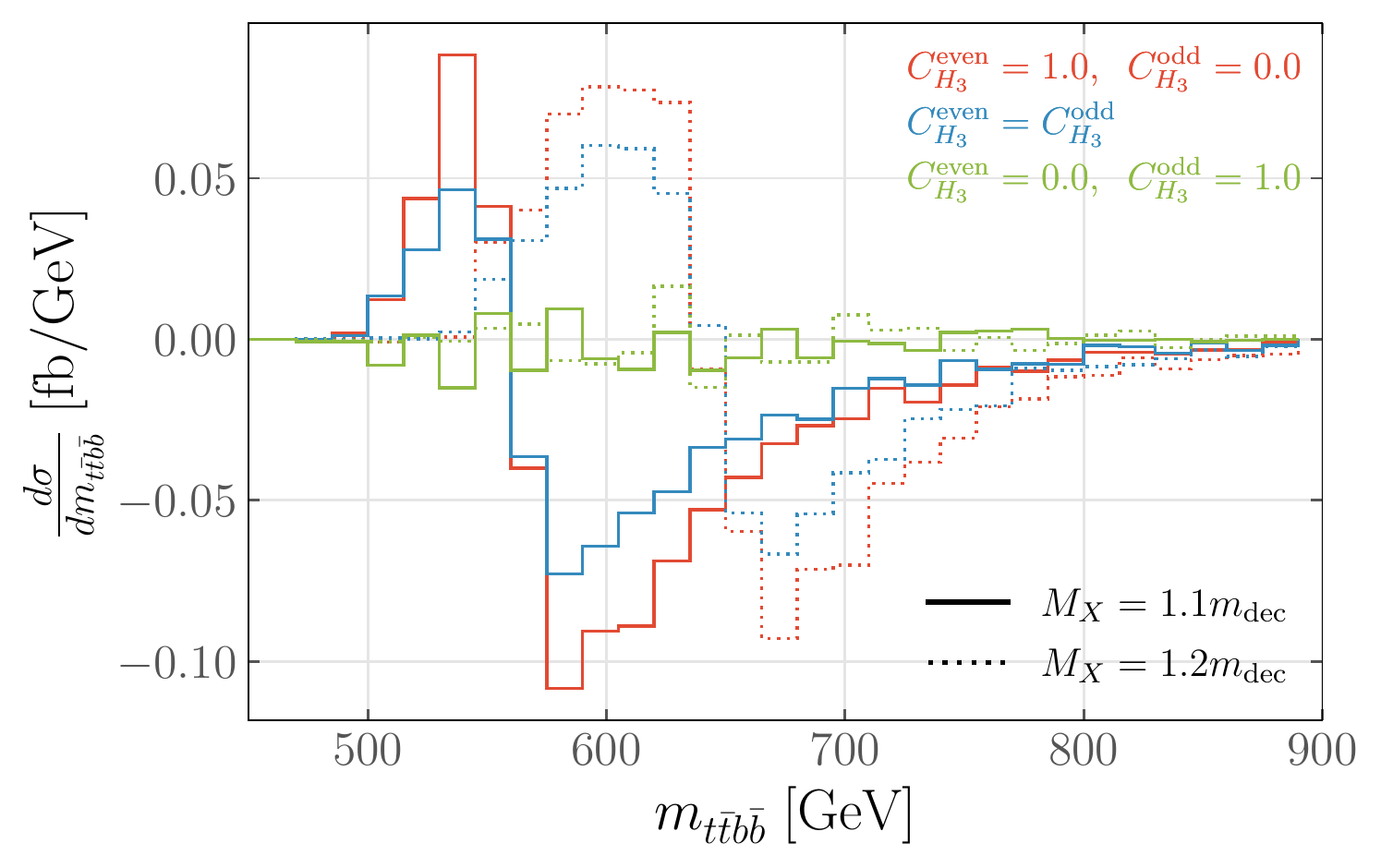}}
   \hskip 0.5cm
	\subfigure[\label{fig_2b}]{\includegraphics[width=0.48\textwidth]{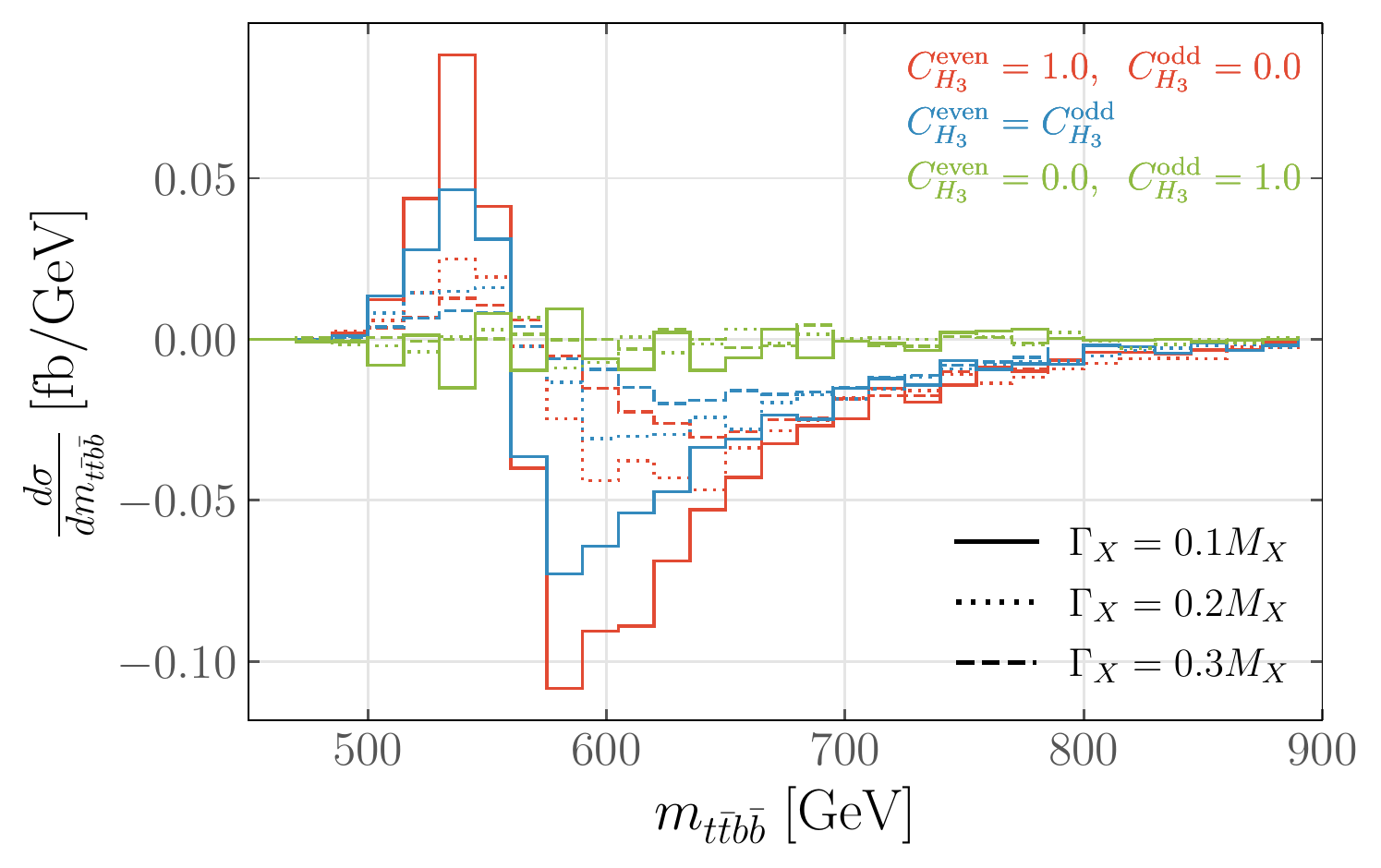}}
   \vskip\baselineskip
	\subfigure[\label{fig_2c}]{\includegraphics[width=0.48\textwidth]{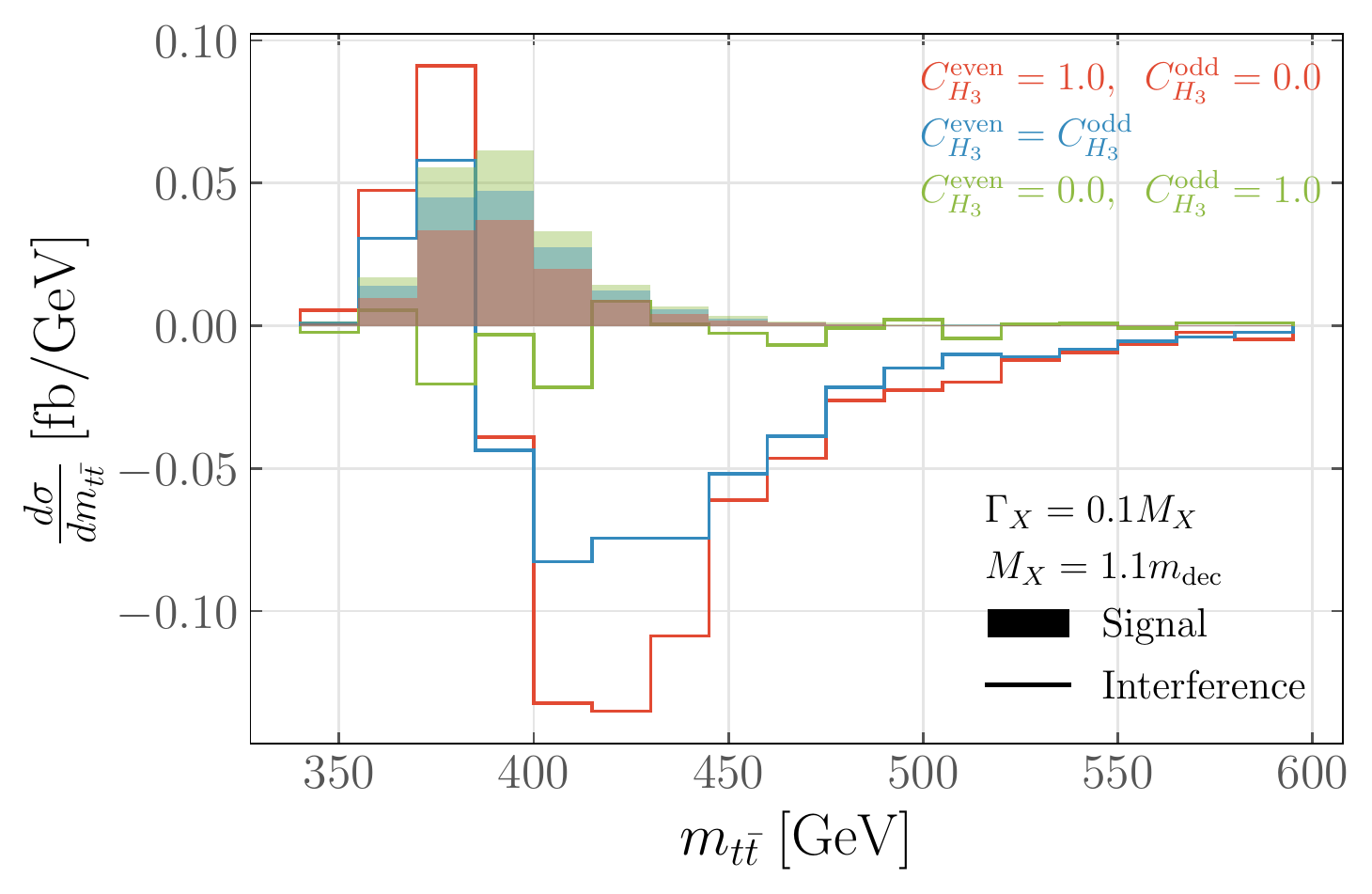}}
   \hskip 0.5cm
	\subfigure[\label{fig_2d}]{\includegraphics[width=0.48\textwidth]{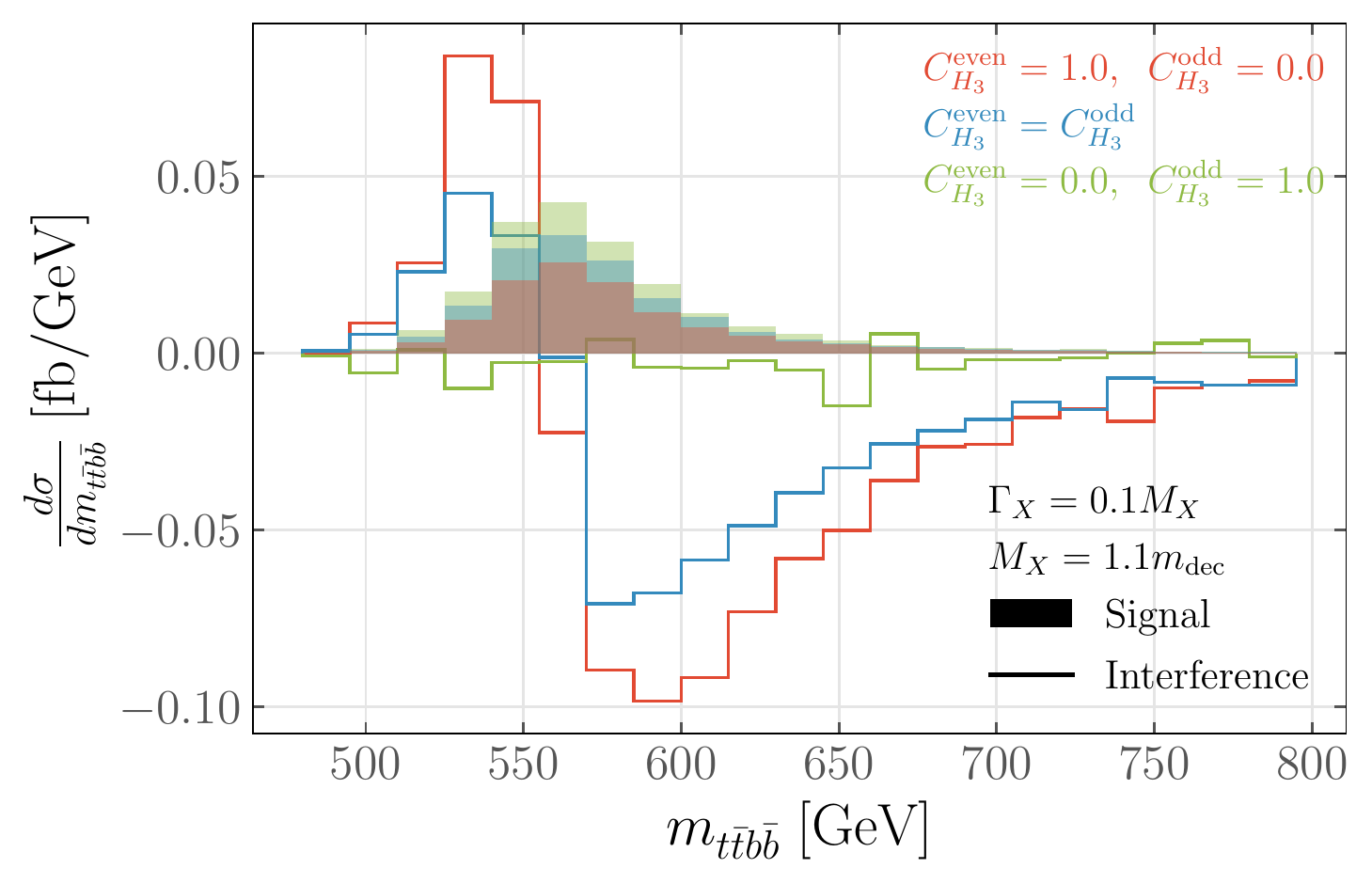}}
   \vskip\baselineskip
	\subfigure[\label{fig_2e}]{\includegraphics[width=0.48\textwidth]{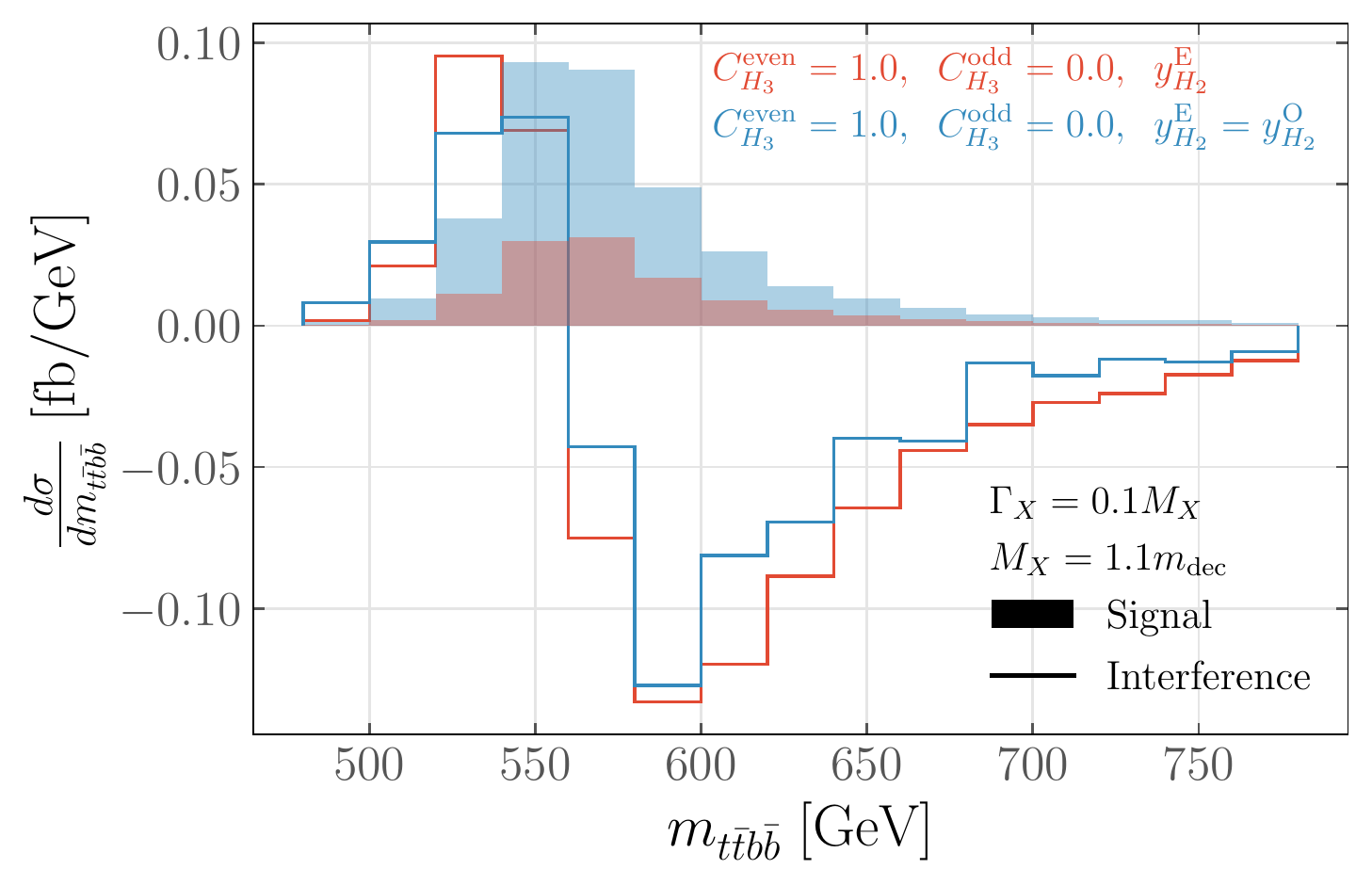}}
   \hskip 0.5cm
	\subfigure[\label{fig_2f}]{\includegraphics[width=0.48\textwidth]{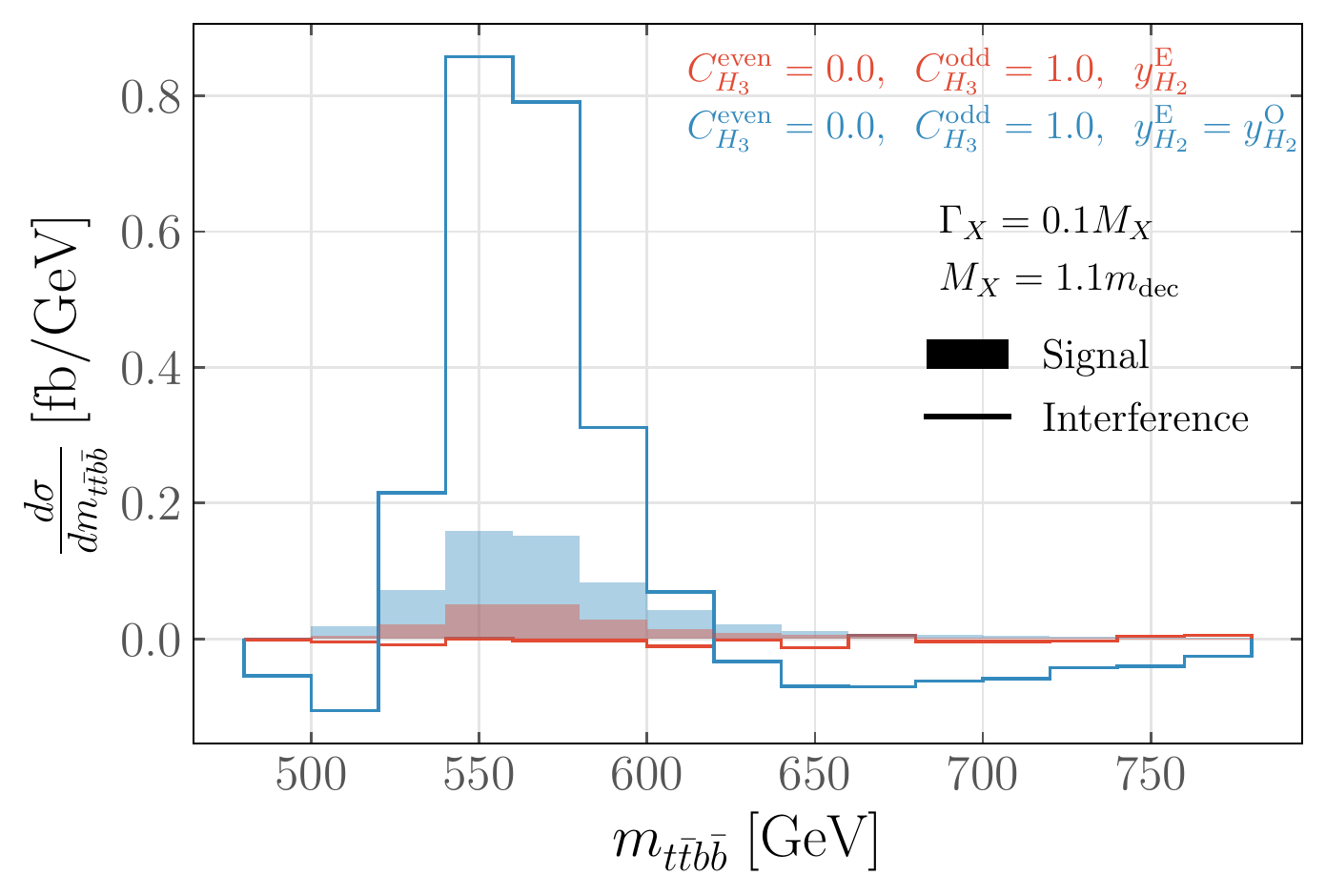}}
	\caption{Figures showing the impact of varying the mass (a) and the width (b) on the SM-BSM interference for different configurations of even and odd coupling phases at the $13$ TeV LHC. $\Gamma_X$ and $M_X$ denote the width and mass respectively, where $X = H_2, H_3$, while $m_{\text{dec}}$ is the sum of the masses of the decay products of the particular scalar. (c) and (d) show the BSM-only contribution along with the interference when reconstructing the $H_2$ and the $H_3$ mass respectively. The aforementioned figures assume that $H_2$ couples to the top only through the CP-even Yukawa coupling $y_{H_2}^E$, while (e) and (f) indicate the enhancement of both signal and interference when $y_{H_2}^E$ and $y_{H_2}^O$ are at a $45^\circ$ angle with unit length, corresponding to an SM-like Yukawa coupling. The $ggH_i$ interactions are computed keeping the full top mass dependence as outlined for Eq.~\eqref{eq:frlagrangian}.}
\label{fig:model_indy_scan}
\end{center}
\end{figure*}

We define the signal cross section as 
\begin{equation}
\hbox{d}\sigma_{S} \sim \hbox{d}{\text{LIPS}}\,|{\cal{M}}_{S}|^2\,,
\end{equation}
where ${\cal{M}}$ is obtained from signal diagrams contributing to topologies of the form of Fig.~\ref{fig:feyndiaga}. Signal-background interference contributions are then given by
\begin{equation}
\hbox{d}\sigma_{I} \sim  2\, \hbox{d}{\text{LIPS}}\, \text{Re}({\cal{M}}_{S} {\cal{M}}_{B}^\ast ) 
\end{equation}
where we include all leading order SM contributions to ${\cal{M}}_B$, which is then dominated by QCD contributions. We keep bottom and top quarks as stable final state objects as their decay is not related to the interference effects that we study in this work. Using their four momentum information we reconstruct the resonance structure of the scalar cascade decay at Monte-Carlo truth level. We simulate $p p \to H_3 \to H_2 h$ events scanning over different configurations of $C_{H_3}^{\text{even}}$ and $C_{H_3}^{\text{odd}}$ while fixing the trilinear $\lam_{H_3 H_2 h}$ and the Yukawa couplings $y_{H_3}$ in such a way that the decays to quarks are SM-like. Additionally, the masses $M_{H_2}$ and $M_{H_3}$ are varied slightly away from the case in which the mass of the resonance is exactly equal to the sum of its decay products. Example histograms for both BSM only and BSM-SM interference are shown in Fig.~\ref{fig:model_indy_scan}.

It can be seen that interference can play a dominant role, depending on the CP violating phase as well as the widths of the scalars. The background is a predominantly CP-even function. Hence, the interference of a CP-odd signal with the background, as a function of a CP-even mass distribution, will show small effects. Defining the decay thresholds $m_{\text{dec}}=2 m_t$ for $H_2$ and $m_\text{dec}=M_{H_2}+M_h$ for $H_3$, we can see in Fig.~\ref{fig_2a} that interference is present for predominantly CP-even couplings of the $H_3$ boson. This holds independently of the separation away from the decay threshold, in the region where we can expect these decays to be significant (i.e. close to the respective $m_{\text{dec}}$) as the background amplitude is a continuous function. The width plays a significant role in how the interference manifests itself, Fig.~\ref{fig_2a}, directly related to the boson propagator structure. We stress again at this point that larger $\Gamma_X/M_X$ values will become sensitive to how the imaginary part of the $H_3$ two-point function is included~\cite{Seymour:1995np,Goria:2011wa,Englert:2015zra}, and only for perturbative values is our choice of a Breit-Wigner propagator justified. 

A crucial question is the impact of the interference when considered in comparison with the resonance peaks $H_{2,3}$, Figs.~\ref{fig_2c} and \ref{fig_2d}. As the SM amplitude does not depend on the BSM spectrum, any interference will necessarily be model-dependent. It is therefore difficult to draw detailed model-independent conclusions (we will revisit this question later using a concrete CP-even scalar extension). For the parameter choices that underpin Fig.~\ref{fig:model_indy_scan} however, we see that, while the $H_3,H_2$ resonance structure is significantly distorted, Figs.~\ref{fig_2c} and~\ref{fig_2d}, their qualitative features are retained. This is similar to the situation in the two-Higgs-doublet model in the $t\bar t$ channel, where the experiments have demonstrated that such effects can be included to set constraints in Refs.~\cite{Aaboud:2017hnm,Sirunyan:2019wph}.

Finally, the relative phase of $H_{2,3}$ is a relevant parameter as it steers interference effects through its decay $H_2\to t\bar t$ and in particular because a dominant CP-even contribution can be obtained from squared CP-odd couplings. As CP-odd couplings are complex this typically reverses the destructive pattern of Figs.~\ref{fig_2c} and \ref{fig_2d}; see Figs.~\ref{fig_2e}, \ref{fig_2f}. This means that an asymmetric cascade could enable an indirect measurement of the CP phases if a discovery is made, through characteristic interference structures, especially when additional information from $t\bar t$ channels is available as discussed in~\cite{Djouadi:2019cbm}.

\section{Cascade interference: Asymmetric decays in the two singlet-extended SM}
\label{sec:tsing}
\subsection{Model and Scan}
In the light of the parton-level results of the previous section, we move on to study a motivated SM extension in order to highlight potential interference effects in concrete scenarios. The extension we focus on is the addition of two real singlet scalar fields, $S_1$ and $S_2$, that give rise to exclusively CP-even BSM signatures~\cite{Ivanov:2017dad}. This scenario has been studied in the literature in various contexts~\cite{Barger:2008jx,Coimbra:2013qq,Choi:2013qra,Ahriche:2013vqa,Costa:2014qga,Costa:2015llh,Ferreira:2016tcu} and can play a role in the strong electroweak phase transition and dark matter~\cite{Lerner:2009xg,Gonderinger:2012rd,Belanger:2012zr,Curtin:2014jma,Kotwal:2016tex,Chiang:2017nmu,Grzadkowski:2018nbc,Cheng:2018ajh}. The model is based on introducing two discrete $\mathbb{Z}_2$ symmetries as
\begin{align}
    \label{eq:TRSM_Z2s}
    \mathbb{Z}_2^{(S_1)}:&\;\;S_1 \to -S_1,\;S_2 \to S_2, \\
    \mathbb{Z}_2^{(S_2)}:&\;\;S_1 \to S_1,\;S_2 \to -S_2,
\end{align}
with the potential given by
\begin{multline}
    \label{eq:TRSM_pot}
   V(H, S_1, S_2) = -\mu_H^2H^\dagger H  - \frac{1}{2}\mu_{S_1}^2S_1^2 -  \frac{1}{2}\mu_{S_2}^2S_2^2  \\
    + \lambda_H(H^\dagger H)^2  + \lambda_{S_1}S_1^4 + \lambda_{S_2}S_2^4\\
     + \lambda_{HS_1}H^\dagger H S_1^2+ \lambda_{HS_2}H^\dagger H S_2^2   
    + \lambda_{S_1S_2}S_1^2S_2^2,
\end{multline}
where all the coefficients are real and $H$ is the Higgs doublet of the SM. After electroweak symmetry breaking (EWSB) each of the scalars acquires a non-zero vacuum expectation value (VEV), $v$, with the VEV of $H$ identified with the SM value of 246 GeV. As in the SM, three of the degrees of freedom of $H$ become the longitudinal polarisations of the $W$ and $Z$ bosons. The non-zero VEVs of $S_{1,2}$ cause the $\mathbb{Z}_2$ symmetries to be broken, yielding a total of three scalar bosons, $\phi_{H,S_1,S_2}$. These scalars undergo mixing, with a transform to the mass eigenstates $H_{1,2,3}$ given by the $3 \times 3$ orthogonal mixing matrix $R$;
\begin{equation}
    \label{eq:TRSM_rot}
    \begin{pmatrix} H_1 \\ H_2 \\ H_3 \end{pmatrix} = R \begin{pmatrix}
    \phi_H \\ \phi_{S_1} \\ \phi_{S_2}
    \end{pmatrix},
\end{equation}
giving three massive CP-even neutral scalar bosons. This mixing matrix can be parametrised with three mixing angles, chosen to be $\theta_{1,2,3}\in[-\pi/2, \pi/2]$, such that
\begin{multline}
R= \\\left(
\begin{matrix}
 c_{\theta_1} c_{\theta_2} & s_{\theta_1} c_{\theta_2} & s_{\theta_2} \\
 -s_{\theta_1} c_{\theta_3}-c_{\theta_1} s_{\theta_2} s_{\theta_3} & c_{\theta_1} c_{\theta_3}-s_{\theta_1} s_{\theta_2} s_{\theta_3} & c_{\theta_2} s_{\theta_3} \\
 s_{\theta_1} s_{\theta_3}-c_{\theta_1} s_{\theta_2} c_{\theta_3} & -s_{\theta_1} s_{\theta_2} c_{\theta_3}- c_{\theta_1} s_{\theta_3} & c_{\theta_2} c_{\theta_3}
\end{matrix}
\right)\,,
\end{multline}
where $s_{\theta_i}=\sin(\theta_i), c_{\theta_i}=\cos(\theta_i)$. Following convention, we take $m_{H_3} \geq m_{H_2} \geq m_{H_1}$ for the mass eigenstates. Additionally we identify the lightest of these bosons with $h$, the observed 125 GeV SM-like Higgs; $H_1 = h$. After requiring a minimum of the extended potential~\cite{Ferreira:2016tcu}, this leaves a total of seven parameters of the model: the two new scalar masses, the VEVs of the two new singlets and the three mixing angles.

\begin{figure*}[!t]
	\begin{center}
		{\includegraphics[width=0.48\textwidth]{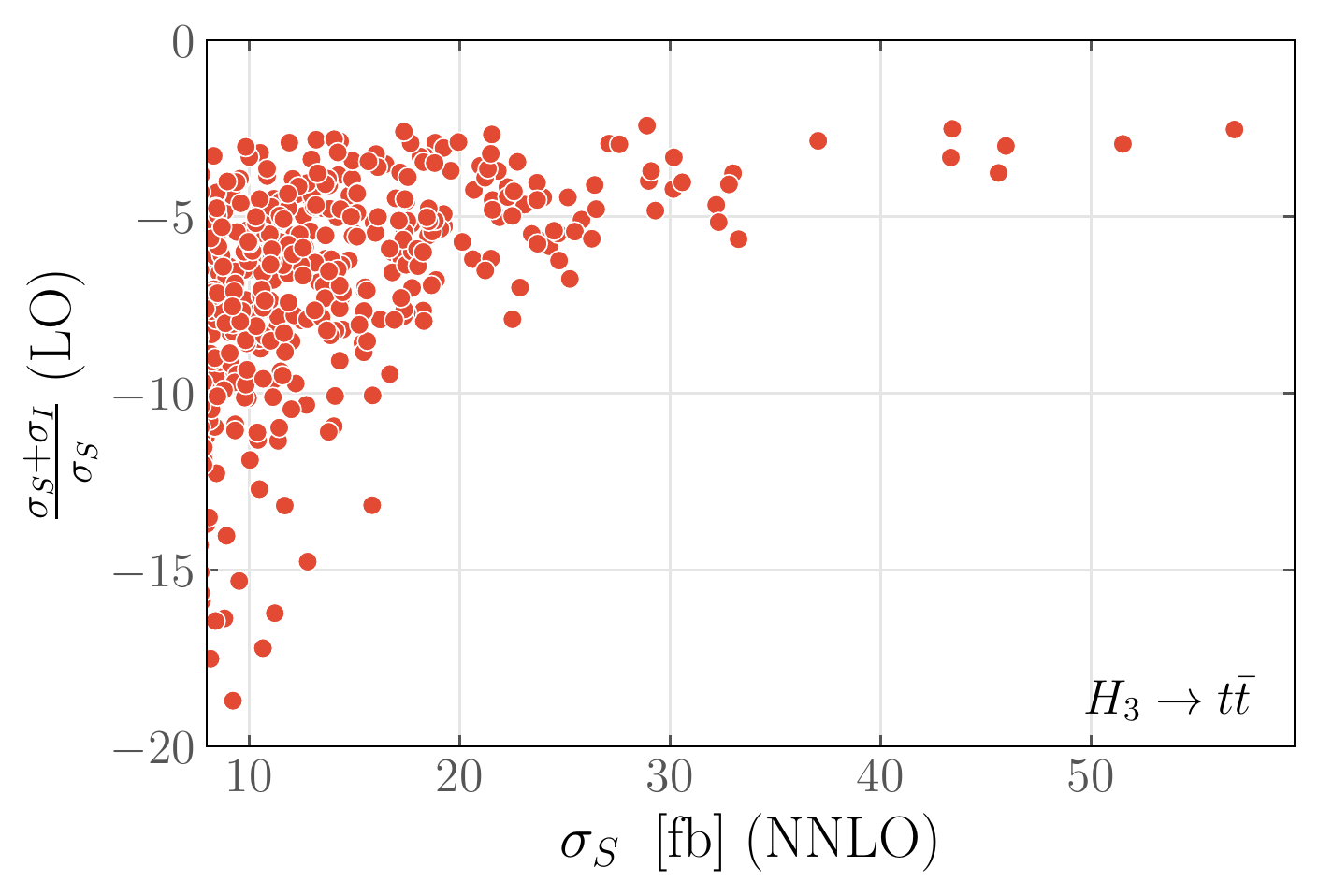}}
    		{\includegraphics[width=0.48\textwidth]{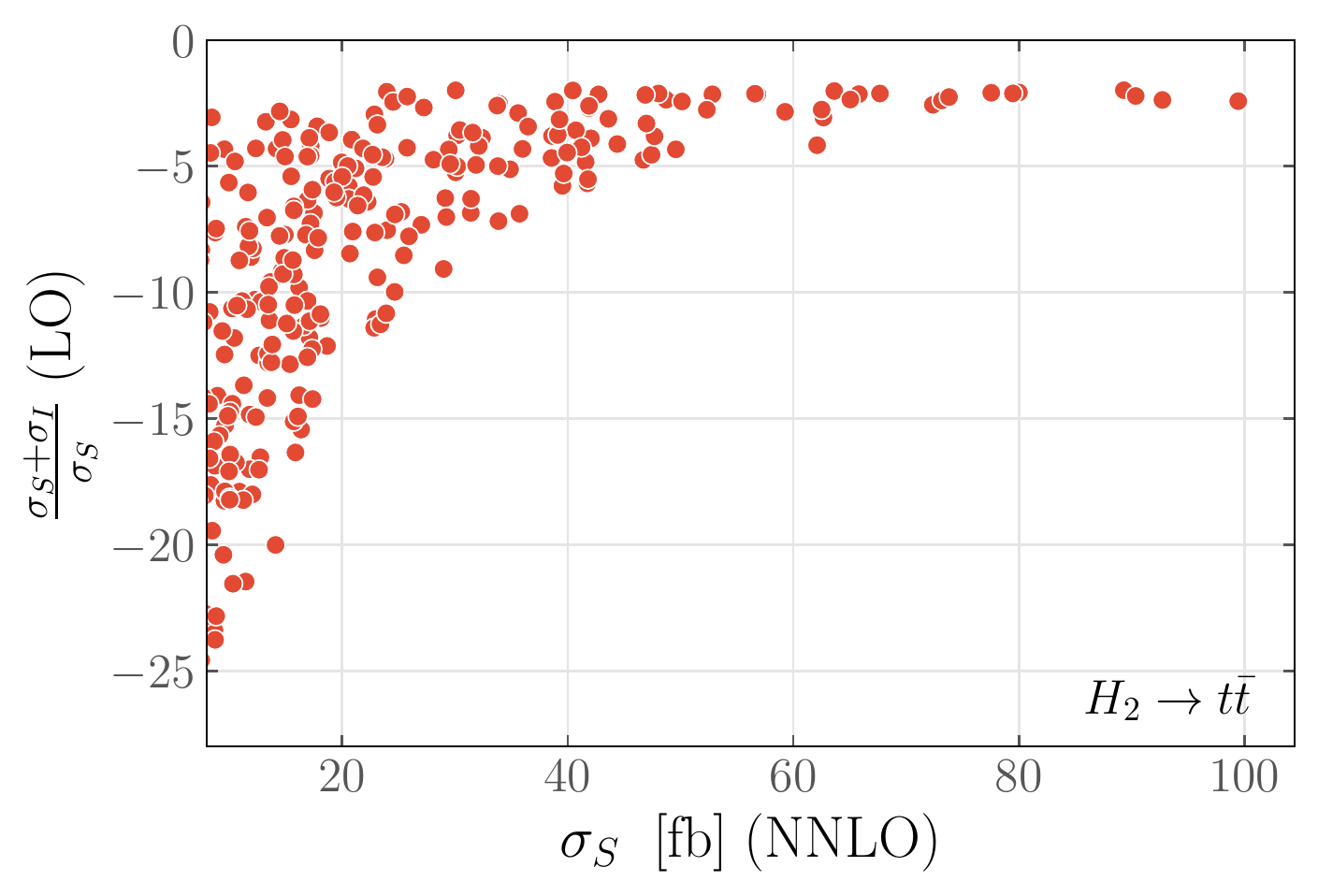}}\\
		{\includegraphics[width=0.48\textwidth]{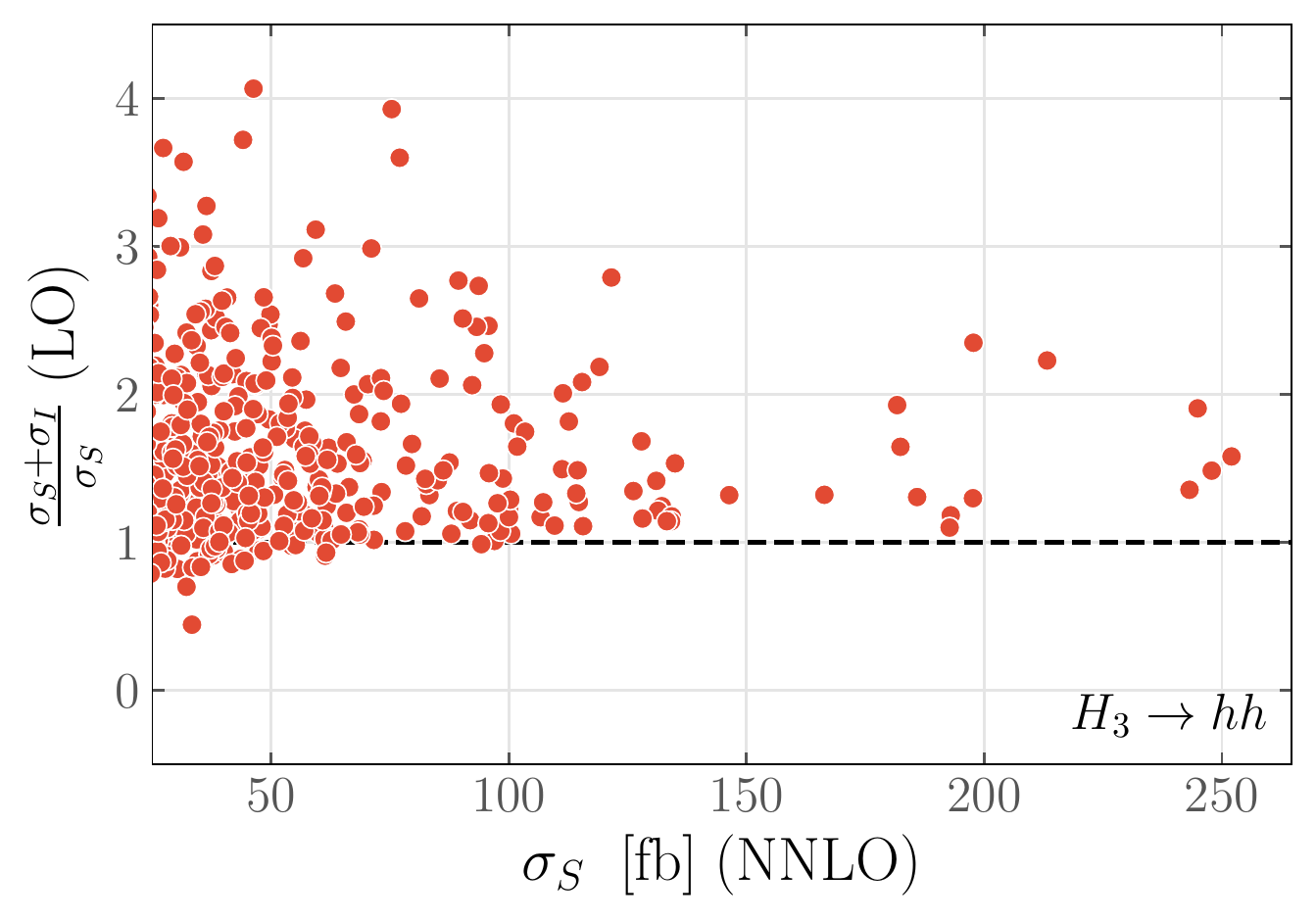}}
    		{\includegraphics[width=0.48\textwidth]{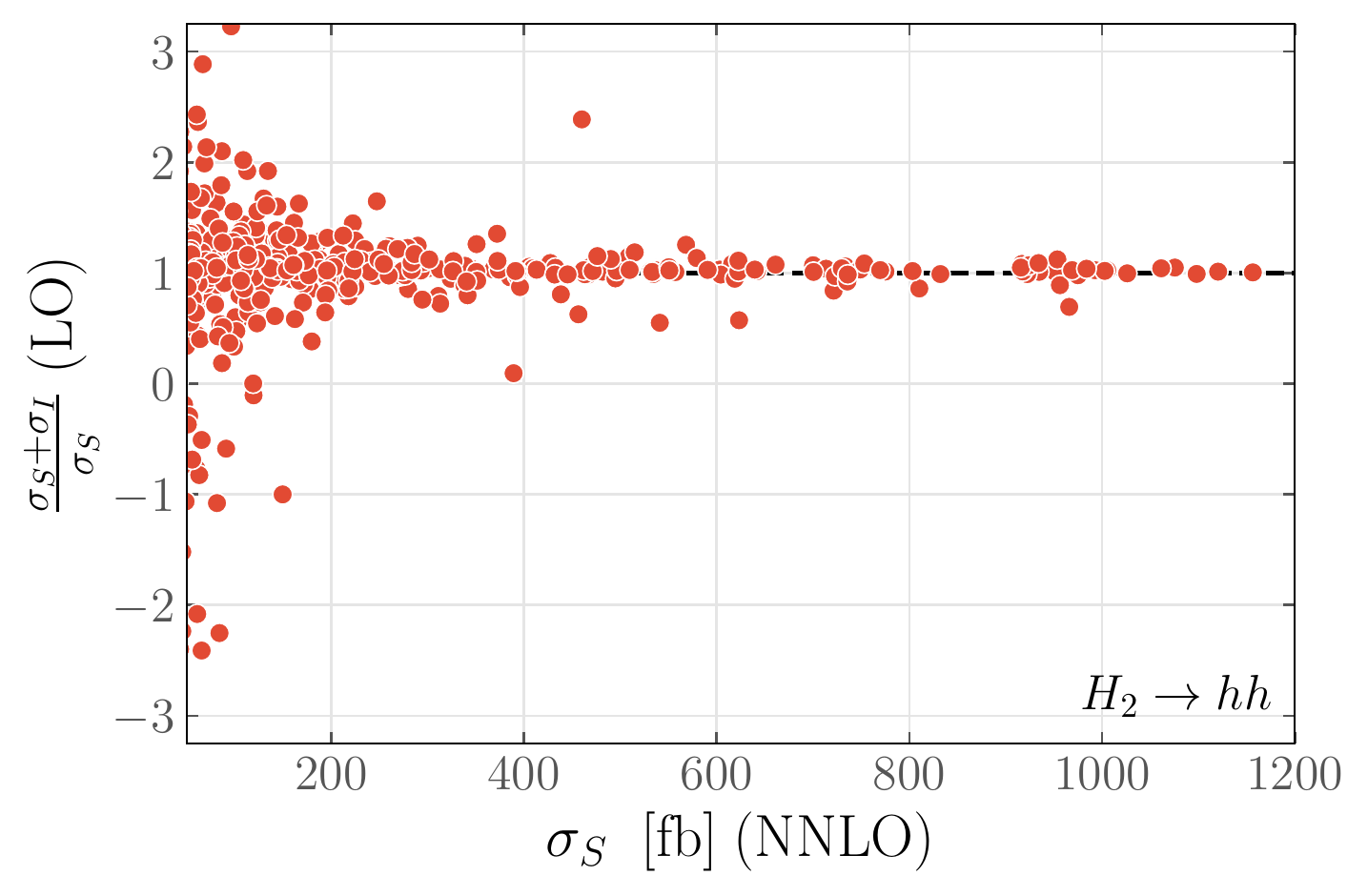}}
		\caption{Scatter plots highlighting the importance of signal-signal interference for resonance search $H_i\to t\bar t,hh$ in the parameter region selected by the scan criteria detailed in the text. 		 \label{fig:standardinf}}
	\end{center}
\end{figure*}

The couplings relevant for Higgs production, decay, and multi-Higgs interactions are then obtained from the overlap of the Higgs bosons with the $\phi_h$ direction. The trilinear Higgs couplings follow from 
\begin{equation}
\parbox{2.3cm}{\includegraphics[width=2.3cm]{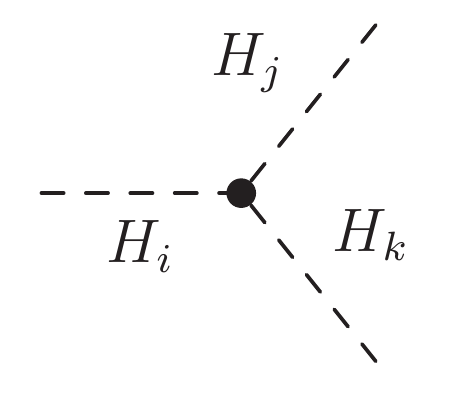}}
=
-i\lambda_{ijk}=iS_{ijk}{\partial^3{\cal{L}}\over \partial H_i \partial H_j \partial H_k}
\end{equation}
where we include a combinatorics factor $S_{ijk}$ to map onto the convention of Eq.~\eqref{eq:frlagrangian}.

To explore the interference patterns for the cascade decay $H_3 \to H_2h \to t\bar{t}b\bar{b}$ we perform a scan over the seven free parameters of the model. Naturally, we require that the cascade channel is open; that $m_{H_3} \geq m_{H_2} + m_h$ and $m_{H_2} \geq 2m_t$, along with requiring that $h$ closely matches the mass and behaviour of the SM Higgs. We achieve the latter through a requirement that the 125 GeV mass eigenstate shows universal coupling modifications $\kappa \gtrsim 0.9$. In parallel, we reflect perturbativity and positivity by enforcing the constraints detailed in~\cite{Kannike:2012pe,Kannike:2016fmd,Robens:2019kga}.

We perform a computation of electroweak oblique observables~\cite{Peskin:1991sw,Peskin:1990zt} and find that the overwhelming majority of points generated complied with the measurements of~Ref.~\cite{Baak:2014ora} after imposing the above selection criteria. Subsequently, we maximise $\sigma_{H_3} \times {\cal B}(H_3 \to H_2h)\geq 4$~fb to identify parameter regions where the asymmetric cascade decay is a priori relevant to discuss the impact of interference effects, with higher-order effects included using the results of the Higgs Cross Section Working group (Refs.~\cite{Spira:1995mt,Heinemeyer:2013tqa,deFlorian:2016spz}). We have computed non-resonant three-body decays of $H_{3,2}$ but find negligible contributions, i.e. the decay of our new states proceeds either like the SM Higgs or through (a)symmetric Higgs cascades, and we compute decays and branching ratios accordingly. A maximised signal cross section along these lines amounts to a situation where $m_{H_3}$ is close to the $m_{H_2}+m_h$ threshold, which leads to a large branching ratio $H_3\to H_2h$.

\subsection{Interference Effects}
Given the scan outlined above, all relevant couplings and masses can be used to assess the interference effects in this particular scenario. Extending the considerations of the previous section, we also include a discussion of interference effects in $pp\to H_{2,3}\to  t\bar t$ and $pp\to H_{2,3}\to h h$ for completeness, including interference that arises from box-like topologies as shown in Fig.~\ref{fig:feyndiagb}, but also from $H_2,h$ $s$-channel contributions as in Fig~\ref{fig:feyndiaga}, with corresponding trilinear couplings contributing to the $t\bar t b\bar b$ final state.\footnote{The analytical and numerical calculations are performed using a combination of \hbox{{\sc{Vbfnlo}}~\cite{Arnold:2008rz,Arnold:2011wj}},~{\sc{FeynArts}},~{\sc{FormCalc}}, and {\sc{LoopTools}}~\cite{Hahn:1999mt,Hahn:2000kx,Hahn:2000jm,Hahn:2001rv}.}

To assess the impact of interference in the different channels, the cross section for each process with and without interference is evaluated within $m_{H_i}^\text{reco} \in \left[ m_{H_i} - 0.15 m_{H_i} , m_{H_i} + 0.15 m_{H_i} \right]$, where the reconstructed heavy scalar mass $m_{H_i}^\text{reco}$ ($i=2,3$) is determined from its decay states. In this particular scenario, signal-background interference in top pair final states is a relevant effect, as can be seen from Fig.~\ref{fig:standardinf}; see also~\cite{Basler:2019nas}. As these results already focus on parameter regions with large asymmetric decay branching fraction, the $H_{3}\to t\bar t$ signal rate is already small, which makes it particularly susceptible to interference effects. However, in the light of expected $t\bar t$ backgrounds, a large interference would not impact the small discovery potential in the first place, when the cross sections are small. Similar to the findings of Ref.~\cite{Basler:2019nas}, the multi-Higgs channels are typically less impacted by signal-signal interference effects, especially when the rate expectation is sizeable. This further motivates the symmetric decays as viable search modes as detailed in Ref.~\cite{Robens:2019kga}. 

\begin{figure}[!t]
	\begin{center}
		{\includegraphics[width=0.48\textwidth]{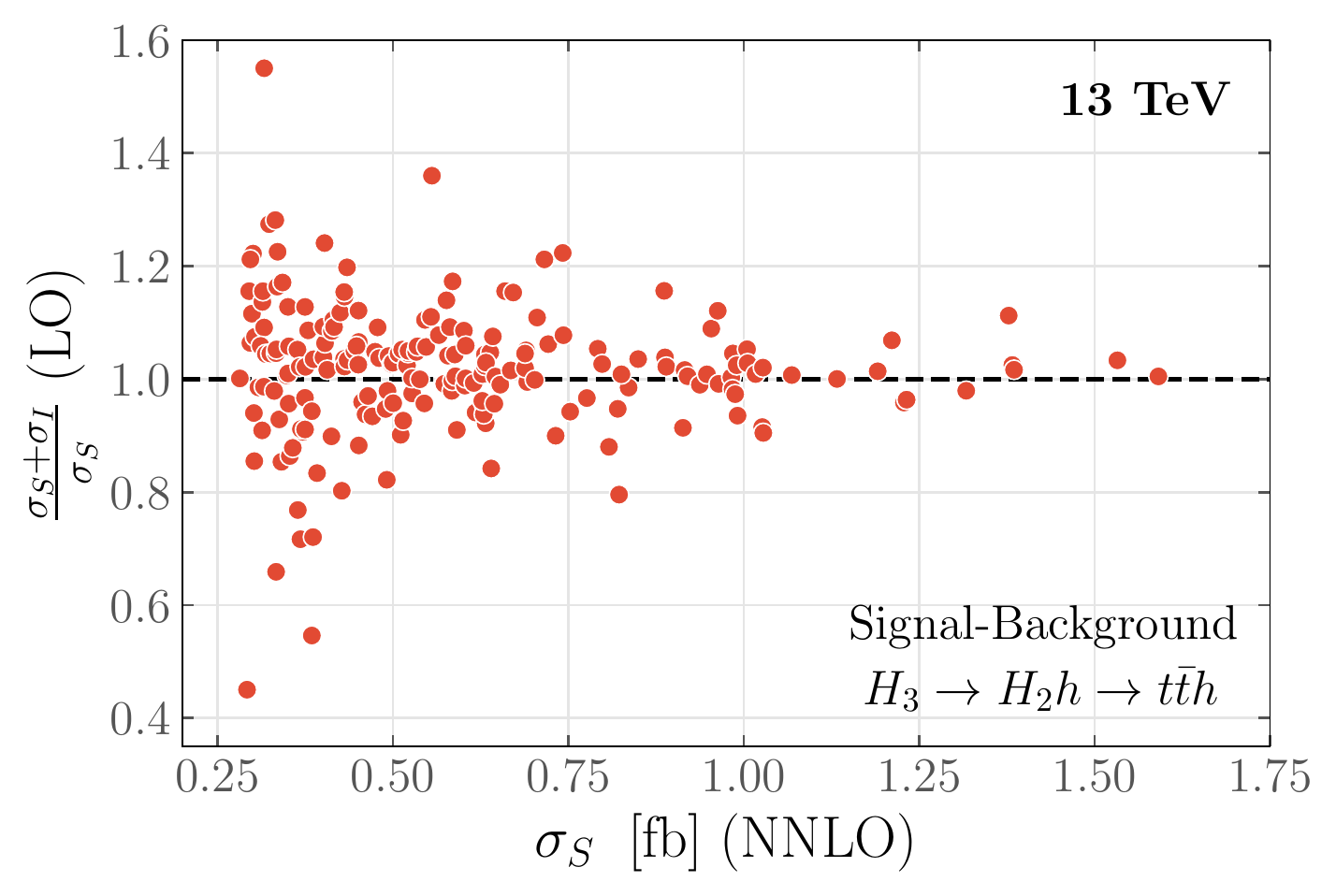}}
		\caption{Cross section of signal only case and impact of signal-background interference in the cascade decay $H_3 \to H_2 h \to t\bar{t}b\bar{b}$ at $13$~TeV.\label{fig:h3_h2h_background}}
		
	\end{center}
\end{figure}

Turning to the asymmetric cascade decays, Fig.~\ref{fig:h3_h2h_background}, we see a similar picture emerging for signal-background interference; it is not a limiting factor in this particular scenario, as the width is small compared to the resonance mass. Signal-signal interference, i.e. interference of the different signal diagrams of Fig.~\ref{fig:feyndiag} can be a more sizeable effect, due to the much busier multi-Higgs phenomenology of this scenario. Such effects need to be taken into account to correctly map out the parameter region of the extended SM if a discovery is made, but also do not impact this particular scenario when the cross sections are large, i.e. when a discovery in this channel becomes more likely.

\begin{figure}[!t]
  \begin{center}
    \includegraphics[width=0.48\textwidth]{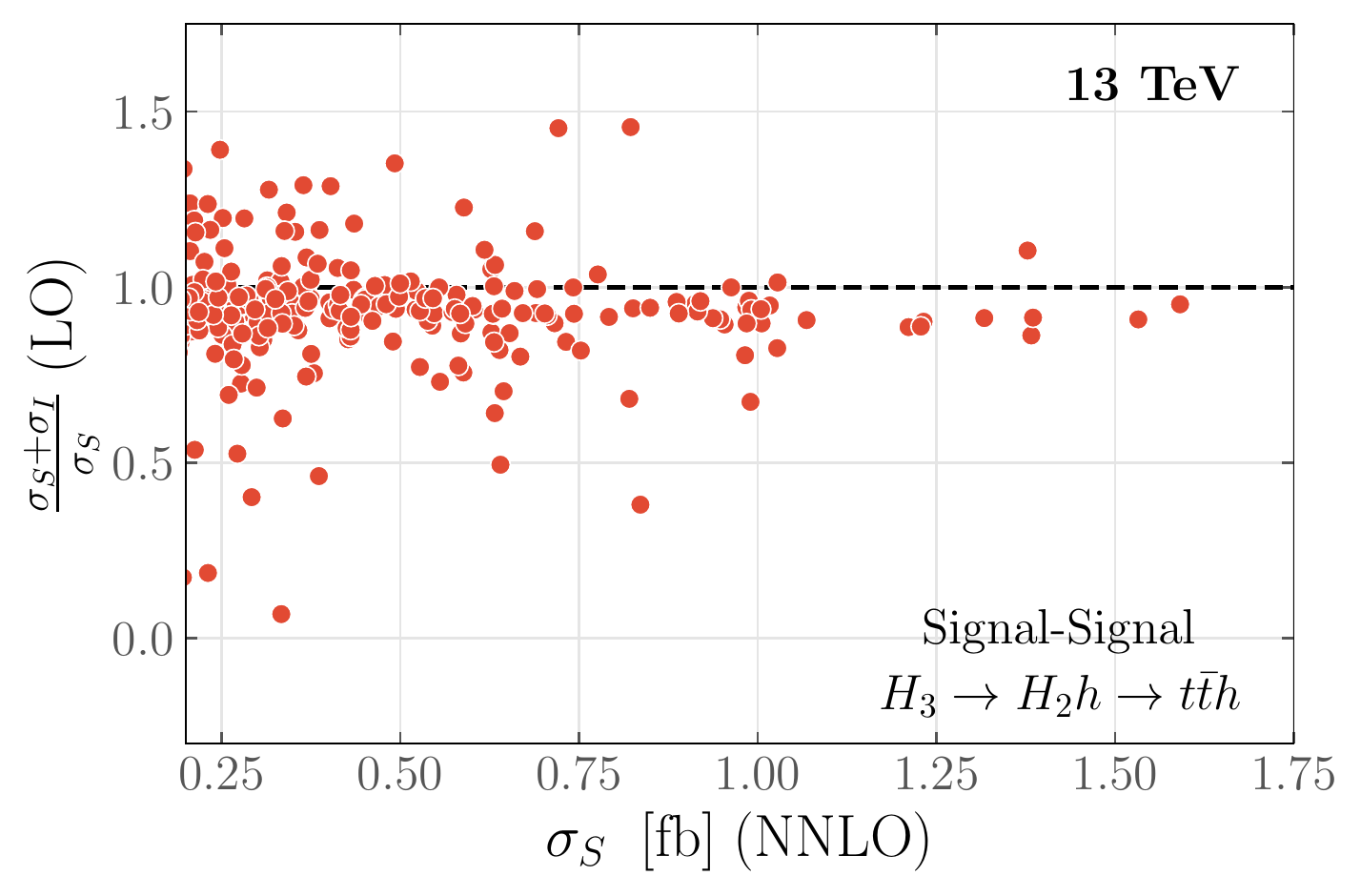}
    \caption{Importance of signal-signal interference for $H_3\to H_2 h$ resonance searches. Included in this comparison are interference contributions with propagating $H_{i\neq 3}$, Fig.~\ref{fig:feyndiaga} as well as non-resonant $gg \to H_2 h$ amplitude contributions that arise from the box topologies of Fig.~\ref{fig:feyndiagb}. \label{fig:h3_h2h_signal}}
	\end{center}
\end{figure}

\begin{figure*}[!t]
	\begin{center}
		{\includegraphics[width=0.48\textwidth]{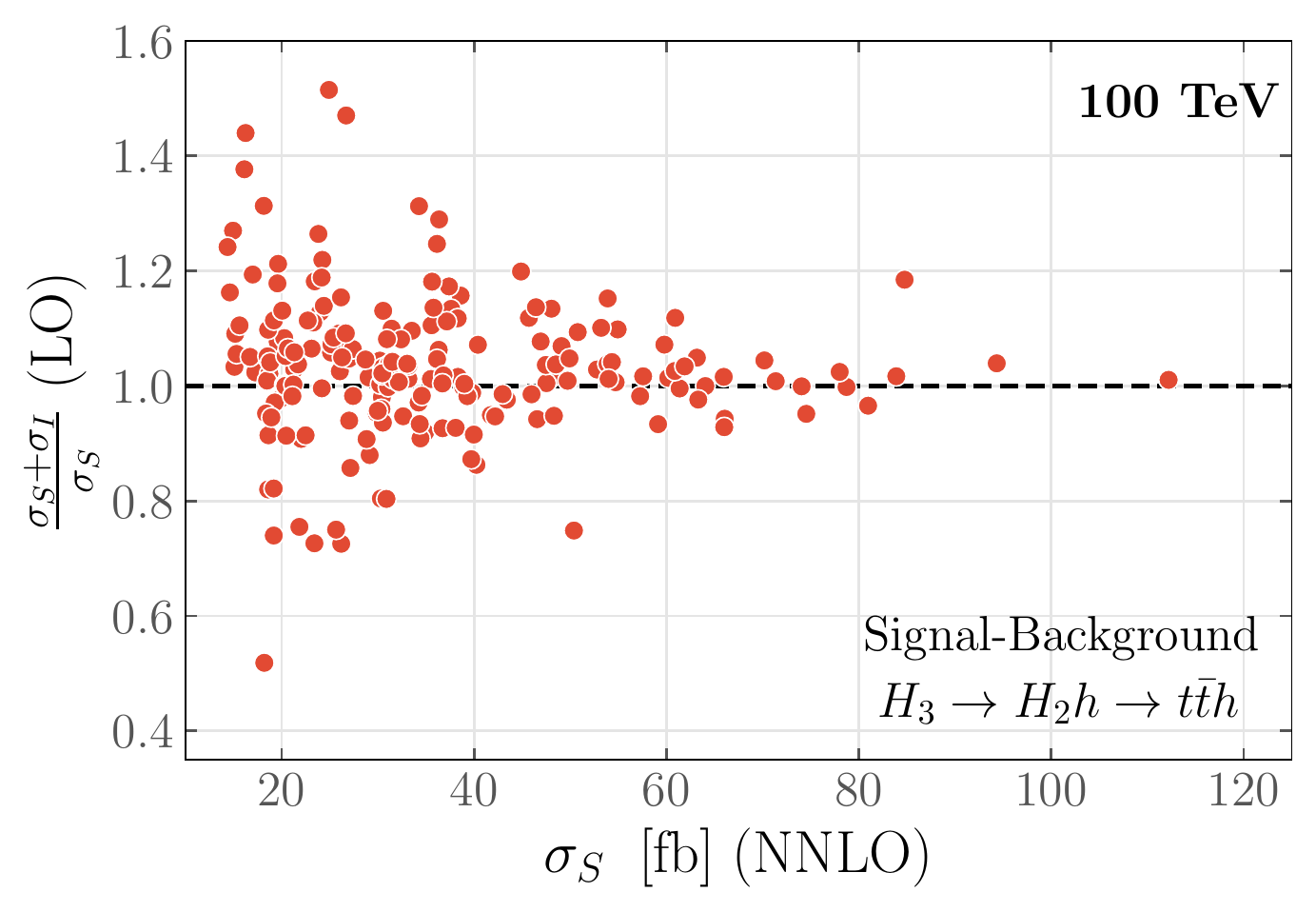}}
		{\includegraphics[width=0.48\textwidth]{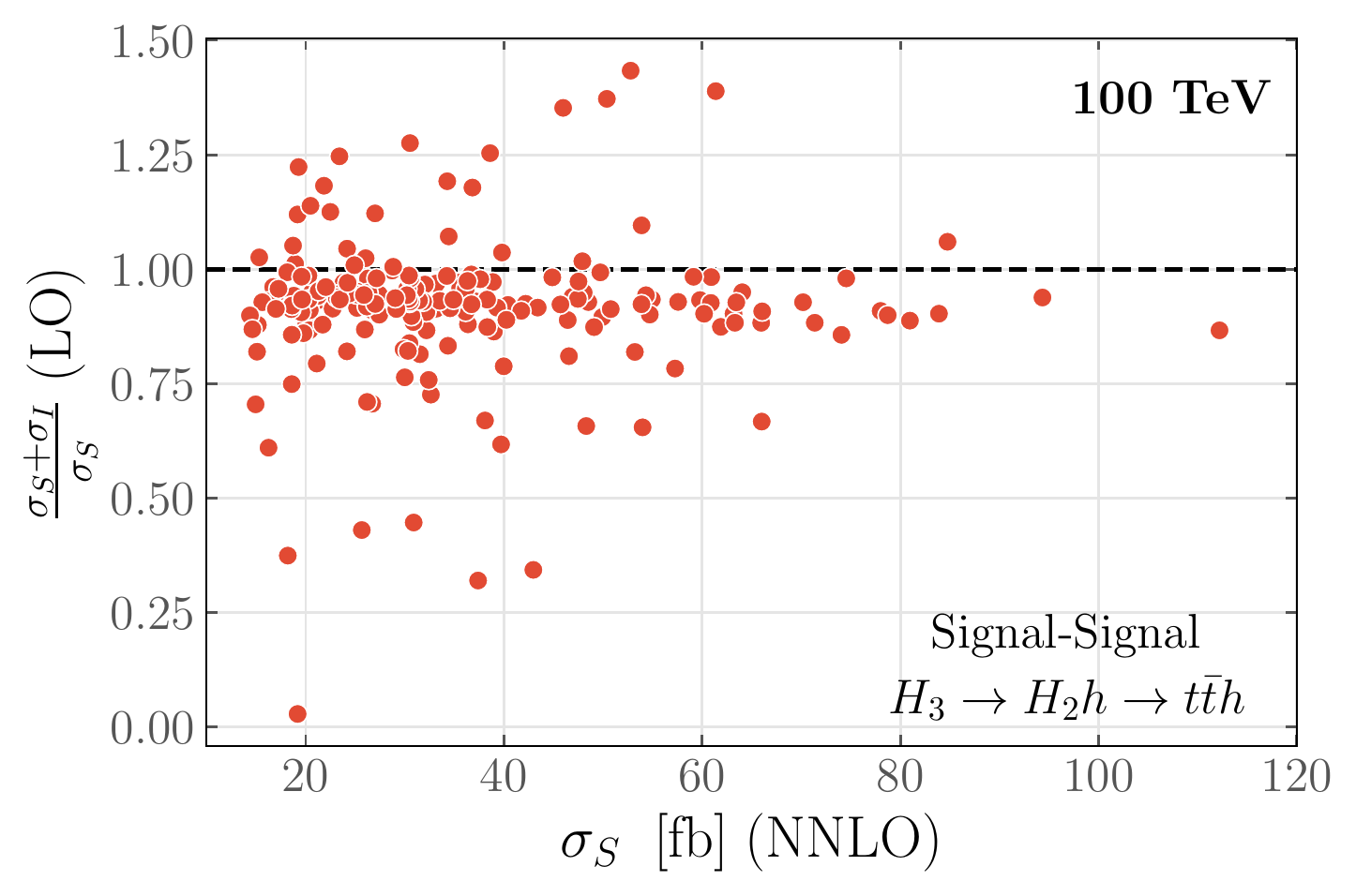}}
		\caption{Similar to Figs.~\ref{fig:h3_h2h_background} and \ref{fig:h3_h2h_signal}, but for an FCC-hh centre-of-mass energy $100$~TeV.\label{fig:h3_h2h_background_100}}
		
	\end{center}
\end{figure*}

Despite the relatively small interference effect from competing contributions, the inclusion of scalar decays to quarks makes this channel phenomenologically challenging to isolate as statistics will be limited at the LHC (see \cite{Englert:2020ntw}). Turning to a future hadron-hadron collider with a centre-of-mass energy of 100 TeV~\cite{Benedikt:2018csr}, as shown in Fig.~\ref{fig:h3_h2h_background_100}, demonstrates that the qualitative features generalise from the LHC to a FCC-hh, with a significantly increased cross section. Therefore, even if a discovery of an asymmetric cascade decay is statistically limited at the LHC, yet a potential discovery in other multi-Higgs channels (see e.g.~Fig.~\ref{fig:standardinf}) further motivates these signatures, the FCC-hh will have the potential to provide additional information without significant impact from signal-background interference.

\subsection{Impact on significance from interference}

In order to assess possible limitations on sensitivity as a result of interference effects in 13 TeV collisions at the HL-LHC, we scan over different values of $M_{H_2}$ and ${M_{H_3}}$. To maximise the HL-LHC potential we focus on machine learning, and recurrent neural networks in particular, which have been shown to provide superior discrimination power for signatures with multiple decay steps~\cite{Englert:2020ntw}.
We fix the branching ratios of the exotic scalars to $\text{BR}(H_3 \to H_2 h) \sim 0.5$ and $\text{BR}(H_2 \to t \bar{t}) \sim 1$ in order to evaluate the trilinear coupling and widths; these choices select parameter regions of our scan where the a priori sensitivity to cascade decays is large. We consider full-showered and decayed events (via $H_2\to t \bar t \to \ell^+ \ell^- b \bar{b} + \slashed{E}_T$, and $h\to bb$) using {\sc{Pythia8}}~\cite{Sjostrand:2014zea}. The showered events are saved in the {\sc{HepMC}} format~\cite{Dobbs:2001ck}. We use {\sc{MadAnalysis}}~\cite{Conte:2012fm,Conte:2014zja,Dumont:2014tja,Conte:2018vmg} to reconstruct the jets using the radius $0.4$ anti-kT algorithm~\cite{Cacciari:2008gp}, implemented in {\sc{FastJet}}~\cite{Cacciari:2011ma,Cacciari:2005hq}. 

The transverse momentum of selected jets must satisfy $p_T(j) > 20$ GeV in a pseudorapidity range $\abs{\eta(j)} < 4.5$. Any candidate b-jet should appear in the central part of the detector to be tagged; $\abs{\eta(b)} < 2.5$, and the b-tagging efficiency is assumed to be 0.8 (see e.g.~\cite{ATLAS:2012ima}). At least four b-jets must be identified for an event to be considered, as well as two isolated leptons\footnote{A lepton is considered isolated if the sum of transverse momenta of jets with a separation from the lepton less than $R = \sqrt{(\Delta \eta)^2 + (\Delta \phi)^2} = 0.3$ is smaller than 20\% of the lepton's $p_T$.} satisfying $p_T(\ell) > 5$ GeV and $\abs{\eta(\ell)} < 2.5$. We also identify the missing transverse momentum as the opposite to the sum of four-momenta of jets and leptons. 

The dominant background contamination for this particular final state arises from $p p \to t \bar{t} b \bar{b}$, while $p p \to t \bar{t} (Z/h \to b \bar{b})$  is also significant. Processes with two gauge bosons ($p p \to b b \bar{b} \bar{b} W^+ W^-$ and $p p \to b b \bar{b} \bar{b} Z Z$) have reduced rates compared to the rest of the processes. The cross sections of $t \bar{t} b \bar{b}$, $t \bar{t} h$ and $t \bar{t} Z$ are rescaled with $K$-factors of $1.8$~\cite{Bredenstein:2009aj}, $1.17$~\cite{deFlorian:2016spz} and $1.2$~\cite{Atlas:2019qfx} respectively.

A neural network implemented with {\sc{Keras}}~\cite{chollet2015keras} is used to classify the events as signal or background and select our final signal region. A Long Short-Term Memory (LSTM) layer of 45 units and tanh activation is used, followed by a dropout with value 0.1 to avoid overfitting. A sigmoid function is used as the recurrent activation of the LSTM. The Adam algorithm~\cite{DBLP:journals/corr/KingmaB14} is used to minimise the categorical cross-entropy and optimise the network, with a 0.001 learning rate and the output layer is activated with the softmax~\cite{10.1007/978-3-642-76153-9_28} function.

\begin{figure*}[!t]
	\begin{center}
		{\includegraphics[width=0.44\textwidth]{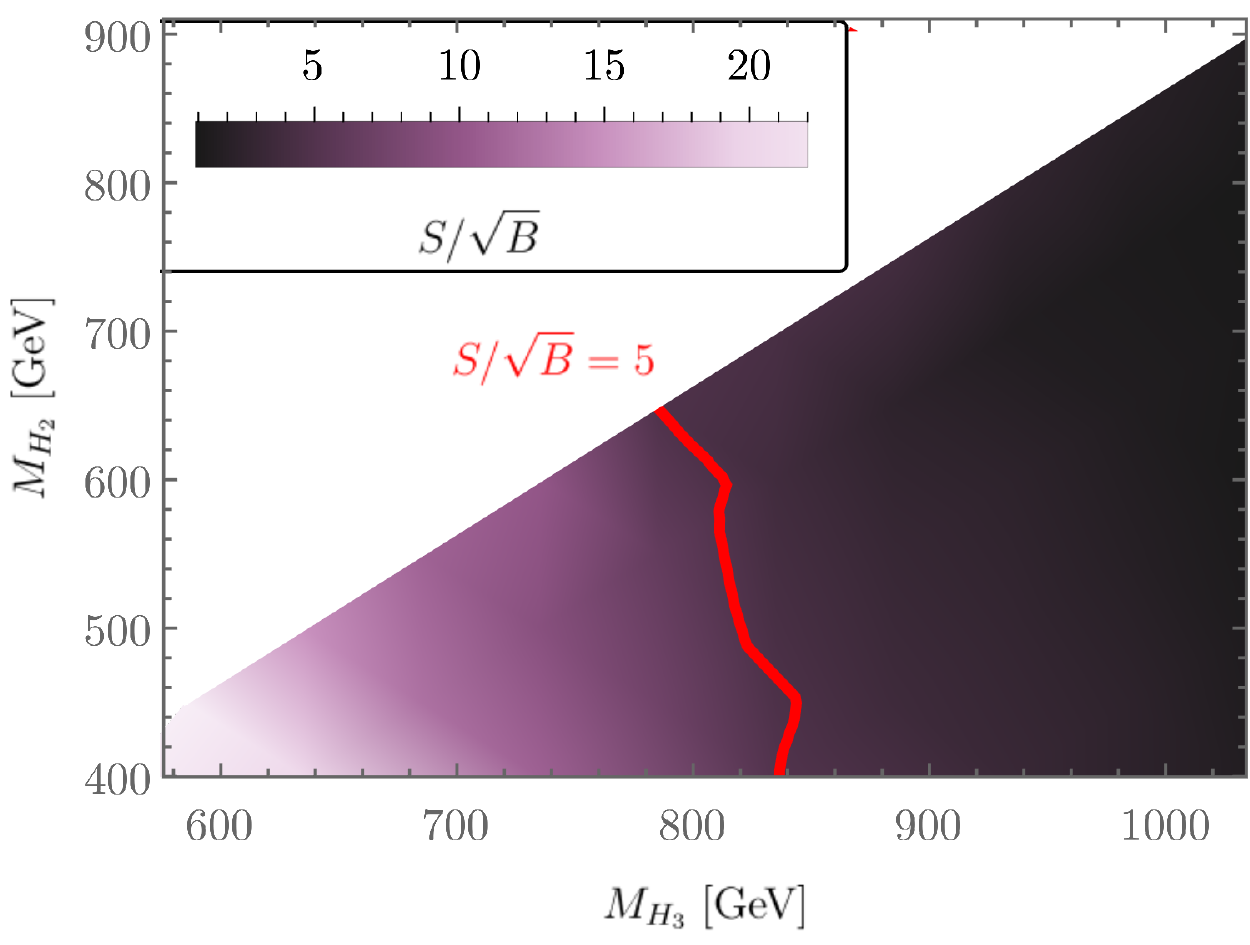}}\hspace{0.4cm}
		{\includegraphics[width=0.44\textwidth]{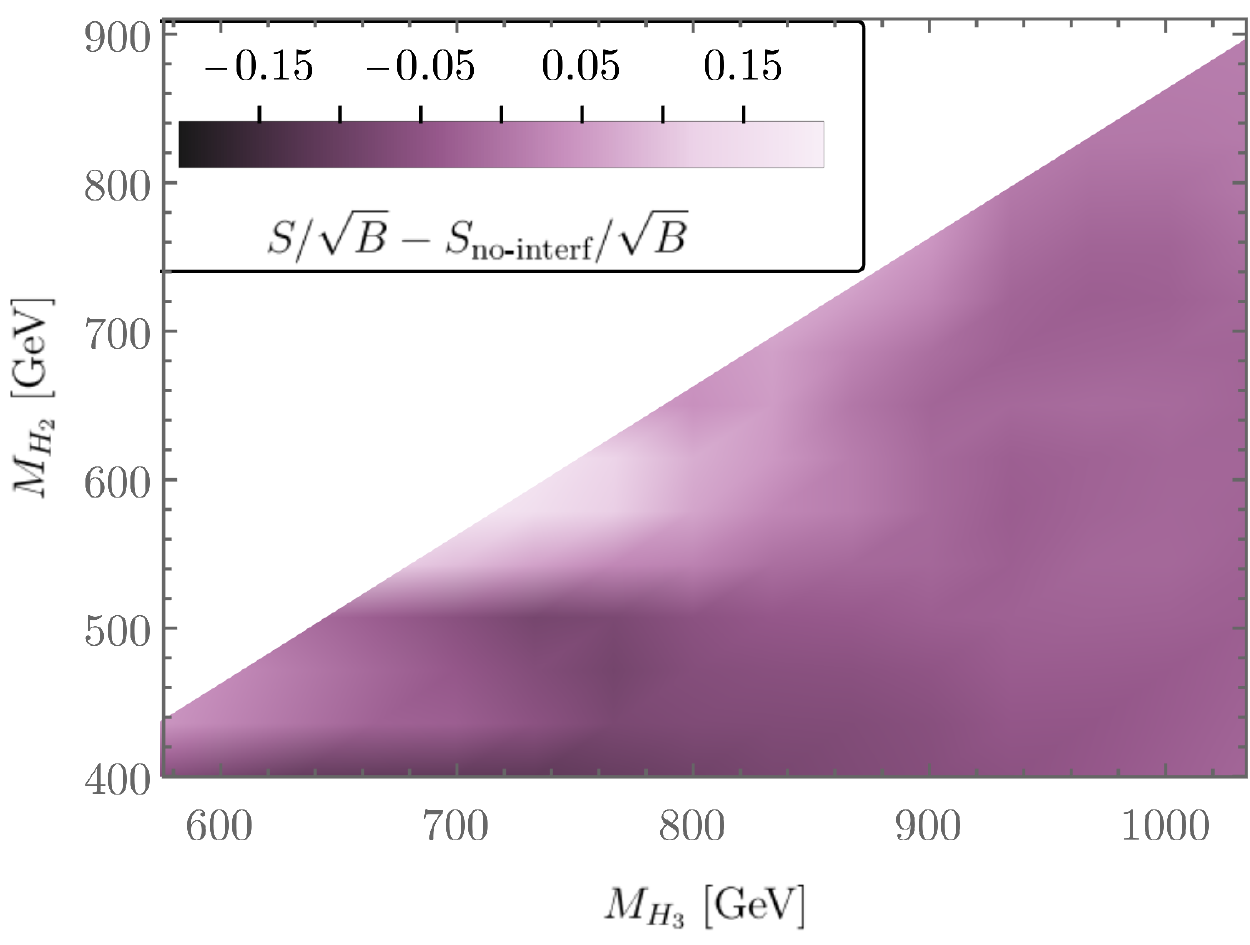}}
		\caption{On the left, the significance $S/\sqrt{B}$ is displayed for an integrated luminosity of 3/ab. The scan over the masses was performed by assuming no CP-odd contributions and fixing the widths and trilinear couplings such that $\text{BR}(H_3 \to H_2 h) \sim 0.5$. The cross sections were normalized to the rates of Ref.~\cite{Englert:2020ntw}. The difference in significance arising from the inclusion of interference is shown on the left, where $S_{\text{no-interf}}$ is the number of events for the signal process without the effects of interference. We do not include theoretical or experimental systematics in this comparison.\label{fig:soverrootb}}
	\end{center}
\end{figure*}

We use a total of 30000 background events
and an equal amount of signal events without interference effects that are split into sets of 81\%, 9\% and 10\% for training, validation and testing, respectively. Our event embedding consists of the missing transverse momentum and the four-momenta of two leptons and four b-jets for each event. We use the energy, transverse momentum, pseudorapidity and azimuthal angle as the basis of our four-momenta and the events are passed to the network in mini-batches of size 200. For each point in the $M_{H_3}$-$M_{H_2}$ plane a network is trained for 100 epochs with an early stopping condition that terminates training if the loss hasn't decreased for ten consecutive epochs. Additionally, the learning rate decays with a factor of 0.1 with a patience of three epochs.  

In order to minimise background systematics, the neural network score threshold is selected for each point in the mass parameter space such that $\sigma_S / \sigma_B$ remains large, where $\sigma_S$ is the signal cross section without interference effects and $\sigma_B$ is the background cross section. The cut on the network score defines our search region and the trained models are used to assess how many positively- and negatively-weighted events pass from a sample that contains only the interference contribution. We define the significance for our study as $S/\sqrt{B}$, where $S$ and $B$ are the total signal and background number of events in the search region at a luminosity of $3/$ab, respectively. We show the significance as well as the change in significance arising from interference in Fig.~\ref{fig:soverrootb}. Our results indicate that the impact of signal-background interference effects on the expected significance that could be achieved at HL-LHC remains small and is not a limiting factor, as also suggested by the parton-level studies of the previous section, in contrast to results in the $p p \to t \bar{t}$ channel. 
The significances are impacted $\lesssim 5\%$ when $\text{BR}(H_3 \to H_2 h)$ is sizeable.
It should also be highlighted that, since the trained network is not specifically trained to classify interference events using any unique characteristics, the small impact on significance arises solely from the fact that interference effects are subdominant and different analysis approaches will yield qualitatively similar results, albeit at potentially smaller significances.


%
%

\section{Summary and Conclusions}
\label{sec:conc}
In this work we considered interference effects in naively well-motivated decay channels of BSM discovery channels as
sensitivity-limiting factors. While such effects can be a tell-tale story of the underlying dynamics if the discovery is not
obstructed~\cite{Carena:2016npr,Djouadi:2019cbm} there is also the possibility of quantum mechanical effects
reducing the sensitivity to a level where the discovery is suppressed. This creates a particularly challenging situation
when such an outcome affects the naively most dominant decay modes, such as searches for $t\bar t $ resonances in 
top-philic extensions of the scalar sector.

In such a situation, we need to turn to more robust final states, where signal-background interference effects are ameliorated. In turn, this then
allows us to fully make use of high precision theoretical calculations for the signal. Naturally, this will involve a smaller discovery
potential in the dominant production modes of such a state. In this work we have considered asymmetric decays of scalar
Higgs sector extensions as candidates for such robust final states. We show, based on a simplified model approach, that interference
effects can be present as a function of the CP properties of the fermion-related interactions of new scalar states. Scanning over a broad
range of simplified model parameters, however, we find that resonance structures are likely to stay intact, in principle enabling an analysis
of the underlying CP structure of the wider scalar interactions if a discovery is made. 

Turning to the more concrete scenario where asymmetric cascade decays arise as a consequence of custodial isospin singlet mixing, we see that interference effects in this purely CP-even setup are largely absent. Tracing our parton-level findings to the fully showered and hadronised final states we see that the significance of detecting this final state in regions of sizeable branching ratio is not limited by signal-background interference. Indeed, using tailored machine learning techniques that unearth the splitting hierarchy of the cascade decay compared to the background will allow us to gain sensitivity to this scenario at the LHC. Owing to the rich scalar interactions, the $H_3\to H_2 h$ channel can become sensitive to signal-signal interference, which then provides a possibility to further explore the new physics scenario if a discovery is made. The machine learning techniques discussed above directly generalise in this instance as the interference contribution is not exploited in the categorisation. This further motivates multi-Higgs and asymmetric cascade decays as signatures to be considered to enhance the BSM discovery potential at the LHC and beyond.

\bigskip
\noindent{\bf{Acknowledgments}} ---
O.A. is supported by a UK Science and Technology Facilities Council (STFC) studentship under grant ST/V506692/1.
C.E. is supported by the STFC under grant ST/T000945/1 and by the IPPP Associateship Scheme.
P.S. is supported by an STFC studentship under grant ST/T506102/1.

\bibliography{references} 

\end{document}